\documentclass[12pt]{article}
\usepackage{amsfonts}
\usepackage{color}

\begin{document}

\def\ba{\begin{eqnarray}}
\def\ea{\end{eqnarray}}
\def\w{\wedge}
\def\g{\gamma}
\def\l{\lambda}

\begin{titlepage}
\title{ \bf Dirac field in topologically massive gravity}
\author{ \"{O}zcan Sert\footnote{\small \tt osert@pau.edu.tr} \hskip 0.3cm
and \hskip 0.1cm Muzaffer Adak\footnote{\small \tt madak@pau.edu.tr} \\ \\
 {\small Department of Physics, Faculty of Arts and Sciences, Pamukkale University}\\
 {\small 20017 Denizli, Turkey } }
 \vskip 1cm
\date{ 26.May.2012 {\it file DiracMielkeBaekler06.tex}}

\maketitle

\begin{abstract}

\noindent We consider a Dirac field coupled minimally to the
Mielke-Baekler model of gravity and investigate cosmological
solutions in three dimensions. We arrive at a family of solutions
which exists even in the limit of vanishing cosmological constant.


\vskip 1cm

\end{abstract}
\end{titlepage}

\section{Introduction}

\noindent Gravity in three dimensions has attracted a lot of
attention since the invention of the topologically massive gravity
by Deser, Jackiw, Tempelton \cite{deser1982}-\cite{nazaroglu2011}.
Another cornerstone is the discovery of the BTZ black hole
solution to the Einstein's theory \cite{banados1992}. Since the
inclusion of torsion makes the gravity models richer, Mielke and
Baekler (MB) generalized the topologically massive gauge model of
gravity by adding a new translational Chern-Simons term to the
lagrangian of the standard topologically massive gravity
\cite{mielke1991}. There is a wide literature on the MB-model of
gravity coupled to various material sources, see for example
\cite{garcia2003}-\cite{blagojevic2009} and the references
therein. We, however, realized that there is a tiny amount of work
considering possible spinor couplings to the Einsteinan gravity
and the standard topologically massive gravity in three
dimensional spacetimes \cite{hehl1971}-\cite{dereli2010}. For the
MB-case the situation is worse \cite{pronin1996}. Thus we intend
to fill in this gap to gain new insights to cosmological problems
in the context of relationship between spinor and gravity.

\medskip

\noindent We are interested in for investigating roles of a Dirac
field in cosmology, because it is essential to a satisfactory
description of relativistic fermions. Furthermore, a
self-interacting or nonlinear Dirac field can yield negative
pressure and thereby accelerate the early and the late-time
expansion of the universe \cite{saha2006}. On the other hand, the
authors show in the Ref.\cite{watanabe2009} that a single Dirac
field can give rise to inflation within the four dimensional
Einstein-Cartan theory, and prove compatibility of the Dirac-field
model with the observations by calculating the power spectrum of
density fluctuations of the Dirac field. Because of the
inflationary (exponential) nature of our solution
(\ref{eq:Solution}) this work may be seen in favor of those
results.

\medskip

\noindent The outline of the paper is as follows. Since we will be
using the coordinate independent algebra of the exterior forms, in
the Section \ref{sec:notation} we fix our notations and
conventions for the Riemann-Cartan spacetimes and a Dirac spinor
in three dimensions. In the Section \ref{sec:fieldeqns} after
introducing the topological gauge lagrangian of gravity and the
Dirac lagrangian we obtain the field equations by varying
independently the total lagrangian with respect to the orthonormal
basis 1-forms, $e^a$, the full connection 1-forms, $\omega^{ab}$
and the adjoint of the Dirac field, $\overline{\Psi}$. In the
Section \ref{sec:solution} firstly we write down the homogenous
and the isotropic metric in the plane polar coordinates and make
torsion ansatz. Secondly we compute the Dirac equation and its
adjoint. This allows us to calculate the Dirac energy-momentum
2-forms explicitly. Then we obtain all field equations via the
computer algebra system, the Reduce and the package Excalc
\cite{hearn1993},\cite{schrufer1994}. Thus we determine a class of
solution for the homogenous and isotropic geometry and a certain
Dirac spinor. We summarize and discuss the results in the Section
\ref{sec:conclusion}.

\section{Notations and Conventions}\label{sec:notation}

\noindent We specify the Riemann-Cartan space-time by a triplet $
\left ( M, g, \nabla \right )$ where $M$ is a 3-dimensional
differentiable manifold equipped with a metric tensor
 \ba
   g = \eta_{ab} e^a \otimes e^b
 \ea
of signature  $(-,+,+)$. $e^a$ is an orthonormal co-frame dual to
the frame vectors $X_a$, that is $e^a(X_b) \equiv \iota_b e^a=
\delta^{a}_{b}$ where $\iota_b := \iota_{X_b}$ denotes the
interior product. A metric compatible connection $\nabla$ can be
specified in terms of connection 1-forms ${\omega^a}_b$ satisfying
$\omega_{ba} = -\omega_{ab}$. Then the Cartan structure equations
 \ba
    de^a + {\omega^a}_b \w e^b = T^a \, , \\
     d{\omega^a}_b + {\omega^a}_c \w {\omega^c}_b = {R^a}_b
 \ea
define the space-time torsion 2-forms $T^a$ and curvature 2-forms
${R^a}_b$, respectively. Here $d$ denotes the exterior derivative
and $\w$ the wedge product. These tensor-valued 2-forms satisfy
the Bianchi identities
 \ba
    DT^a = R^a{}_b \w e^b \; , \quad \quad DR^a{}_b =0 \, .
 \ea
Because of the first Bianchi identity, the Ricci tensor does not
need to be symmetric in the Riemann-Cartan spacetimes. We fix the
orientation of space-time by choosing the volume 3-form ${}^*1 =
e^0 \wedge e^1 \wedge e^3$ where ${}^*$ is the Hodge star map. In
three dimensional space-times with Lorentz signature for any
$p$-form ${}^{**}=-1$. We will use the abbreviations $e^{ab
\cdots} := e^a \w e^b \w \cdots $ and $\iota_{ab\cdots} := \iota_a
\iota_b \cdots$. It is possible to decompose the connection
1-forms in a unique way as
 \ba
   {\omega^a}_b = \hat{\omega}^a_{\;\;b} +{K^a}_b
   \label{eq:FullConnec}
 \ea
where $\hat{\omega}^a_{\;\;b}$ are the zero-torsion Levi-Civita connection 1-forms satisfying
 \ba
     de^a + \hat{\omega}^a_{\;\;b} \w e^b = 0
 \ea
and ${K^a}_b$ are the contortion 1-forms satisfying
 \ba
    {K^a}_b \w e^b = T^a \, . \label{eq:Contor1}
 \ea
The curvature 2-forms are also decomposed in a similar way:
 \ba
    {R^a}_b = \hat{R}^a_{\;\;b} + \hat{D}{K^a}_b + {K^a}_c \w {K^c}_b
 \ea
with
$$
\hat{D}{K^a}_b = d{K^a}_b + \hat{\omega}^a_{\;\;c} \w {K^c}_b - \hat{\omega}^c_{\;\;b} \w {K^a}_c \,.
$$

\medskip

\noindent We are using the formalism of Clifford algebra $\mathcal{C}\ell_{1,2}$-valued exterior forms.
$\mathcal{C}\ell_{1,2}$ algebra is generated by the relation among the orthonormal basis
$\{\gamma_0,\gamma_1,\gamma_2\}$
 \begin{equation}
    \gamma^a \gamma^b + \gamma^b \gamma^a = 2\eta^{ab} \, .
 \end{equation}
One particular representation of the $\gamma^\alpha$'s is given by the following Dirac matrices
 \begin{eqnarray}
   \gamma_0 = \left(\begin{array}{cc}
                0 & 1 \\
                -1 & 0
              \end{array}\right) \, , \;
   \gamma_1 = \left(\begin{array}{cc}
                0 & 1 \\
                1 & 0
              \end{array}\right) \, , \;
   \gamma_2 = \left(\begin{array}{cc}
                1 & 0 \\
                0 & -1
              \end{array}\right) \, .
 \end{eqnarray}
In this case a Dirac spinor $\Psi$ can be represented by a 2-component column matrix.
Thus we write explicitly the covariant exterior derivative of $\Psi$, its Dirac conjugate and the curvature of the spinor bundle, respectively,
 \ba
  D\Psi = d\Psi + \frac{1}{2}  \sigma_{ab} \Psi \omega^{ab} \; , \quad
  D\overline{\Psi} = d\overline{\Psi} - \frac{1}{2} \overline{\Psi} \sigma_{ab}  \omega^{ab} \; , \quad
  D^2\Psi = \frac{1}{2}R^{ab} \sigma_{ab} \Psi
 \ea
where $\sigma_{ab}:= \frac{1}{4}[\gamma_a , \gamma_b]=\frac{1}{2}
\epsilon_{abc} \gamma^c$ are the generators of the Lorentz group
$SO(1,2)$. The Dirac adjoint is $\overline{\Psi}:= \Psi^\dag
\gamma_0$. We frequently make use of the identity
 \ba
     \gamma_c \sigma_{ab} + \sigma_{ab} \gamma_c = \epsilon_{abc} \, .
 \ea

\section{Field Equations} \label{sec:fieldeqns}

\noindent The field equations of our model are obtained by varying the action
 \ba
   I[e^a,\omega^{ab},\overline{\Psi}] = \int_M \left ( L_G + L_D  \right )
 \ea
where the gravitational lagrangian density 3-form is given by
Mielke and Baekler \cite{mielke1991}
 \ba
  L_G = \frac{a}{2} R_{ab} \w {}^*e^{ab} + \lambda {}^*1
  + \frac{b}{2} T^a \w e_a + \frac  {c}{2}({\omega^a}_b \wedge d {\omega^b}_a
  + \frac{2}{3}{\omega^a}_b \wedge {\omega^b}_c \wedge {\omega^c}_a )
 \ea
and the (hermitian) Dirac lagrangian density 3-form
 \ba
   L_D = \frac{i}{2} \left ( \overline{\Psi}\, {}^*\gamma \w D \Psi - D\overline{\Psi} \w \, {}^*\gamma \Psi \right )
   + i m \overline{\Psi} \Psi \, {}^*1 \, . \label{eq:DirLag}
 \ea
Here the gravitational constants $a$, mass $m$ and the Dirac field
$\Psi$ have the dimension of $length^{-1}$,  the gravitational
constant $b$ has the dimension of $length^{-2}$, $c$ is
dimensionless constant, and the cosmological constant $\lambda$
has the dimension of $length^{-3}$. The case $b=0$ corresponds the
topologically massive gravity with cosmological constant. The
hermiticity of the lagrangian (\ref{eq:DirLag}) leads to a charge
current which admits the usual probabilistic interpretation.
$\frac{b}{2} T^a \w e_a$ is known as the {\it torsional
Chern-Simons term} which corresponds the usual Chern-Simons 3-form
for the curvature, $ (1/2) ({\omega^a}_b \w d{\omega^b}_a + (2/3)
{\omega^a}_b \w {\omega^b}_c \w {\omega^c}_a)$, for the curvature.

\medskip

\noindent We obtain the field equations via independent variations with
respect to $e^a, \omega^{ab}, \overline{\Psi}$. Thus
$e^a$-variation yields the FIRST equation
 \ba
    -\frac{a}{2} \epsilon_{abc} R^{bc} - \lambda {}^*e_a  - bT_a
= \tau_a \, , \label{eq:FldEqCoframe}
 \ea
$\omega^{ab}$-variation yields the SECOND equation
  \ba
    -\frac{a}{2} \epsilon_{abc} T^c  + \frac{b}{2} e_{ab}  + cR_{ab} = \Sigma_{ab}  \, , \label{eq:FldEqConnec}
 \ea
and $\overline{\Psi}$-variation yields the Dirac equation
 \ba
     {}^*\gamma \w (D-\frac{1}{2}\mathcal{V})\Psi + m \Psi \, {}^*1
     =0 \, , \label{eq:DirEqn}
 \ea
where $ \Sigma_{ab}= - \frac{i}{4} \overline{\Psi} \Psi e_{ab}$ is
the Dirac spin angular momentum 2-form and $\tau_a$ is the Dirac
energy-momentum 2-form
 \ba
  \tau_a = \frac{i}{2}  {}^*e_{ba} \w \left[ \overline{\Psi} \gamma^b
  (D\Psi) - (D\overline{\Psi}) \gamma^b \Psi \right] + im\overline{\Psi} \Psi
  {}^*e_a \, .
 \ea
For future convenience by using the Dirac equation
(\ref{eq:DirEqn}) and its conjugate
$(D-\frac{1}{2}\mathcal{V})\overline{\Psi} \w {}^*\gamma -m
\overline{\Psi} {}^*1=0$ we rewrite the Dirac energy-momentum and
the spin angular momentum 2-forms as
 \ba
    \tau_a &=& -\frac{i}{2} \left[ \overline{\Psi} \gamma_b (\partial_a \Psi) - (\partial_a\overline{\Psi}) \gamma_b \Psi
    \right] {}^*e^b + \mathcal{S} \omega_{bc,a} e^{bc}
    \label{eq:DirEnMom}\\
 \Sigma_{ab}&=& - \mathcal{S} e_{ab} \label{eq:DirSpin}
 \ea
where $\partial_a := \iota_a d$, $\omega_{bc,a} := \iota_a
\omega_{bc}$ and $\mathcal{S} := \frac{i}{4} \overline{\Psi}
\Psi$.

\medskip

\section{A Class of Cosmological Solution} \label{sec:solution}

\noindent We consider the homogeneous and isotropic metric in the plane polar coordinates $(t, r, \phi)$
 \ba
    g = -  dt^2 + \frac{Q^2(t)}{1-kr^2} dr^2 + r^2 Q^2(t) d \phi^2
 \ea
with the expansion factor $Q(t)$ and curvature index $k$.
We  choose  the orthonormal basis 1-forms
 \ba
   e^0 =  dt \, , \quad e^1 =\frac{ Q(t)}{\sqrt{1-kr^2}} dr   \, , \quad e^2 = r Q(t) d \phi ,
 \ea
leading to the Levi-Civita connection 1-forms
 \ba
    \hat{\omega}^{0}_{\; \;1} = \frac{\dot{Q}}{Q} e^1 \, , \quad
    \hat{\omega}^{0}_{\; \; 2} = \frac{\dot{Q}}{Q} e^2 \, , \quad
    \hat{\omega}^{1}_{\; \; 2} = - \frac{\sqrt{1-kr^2}}{rQ}  e^2
 \ea
where dot denotes the derivative with respect to $t$. Two
independent functions are enough to describe the most general
torsion preserving  homogeneity and isotropy of the spacetime
\cite{goenner1984}. Thus we choose the torsion
 \ba
  T^0 = u(t)e^{01}+ v(t) e^{12} \, , \quad T^1 = v(t)e^{02} \, , \quad T^2 =
-v(t)e^{01} -  u(t) e^{12} \, . \label{eq:tors}
 \ea
We then calculate the contortion 1-forms via (\ref{eq:Contor1})
 \ba
    K^{0}_{\; \;1} =  ue^{0}- \frac{v}{2} e^{2} \, , \quad
    K^{0}_{\; \; 2} = \frac{v}{2}e^1\, , \quad
    K^{1}_{\; \; 2} = -ue^{2}+ \frac{v}{2} e^{0} \, .
    \label{eq:contor2}
 \ea
Now we write explicitly the full connection 1-forms with the help
of the equation (\ref{eq:FullConnec})
 \ba
    \omega_{01} = -ue^0  -\frac{\dot{Q}}{Q}e^{1}   + \frac{v}{2} e^{2} \, , \quad \quad
       \omega_{02} = - \frac{v}{2} e^1 - \frac{\dot{Q}}{Q}e^2 \, , \nonumber \\
    \omega_{12} = \frac{v}{2} e^{0} - \left( u + \frac{\sqrt{1-kr^2}}{rQ} \right) e^2 \, .
 \ea
Thus the curvature 2-forms are computed as
 \ba
   & &{R^0}_1 = \frac{ (4\ddot{Q} +  v^2Q)e^{01}-2\dot{v}Qe^{02} + 2uvQe^{12} }{4Q},       \\
   & &{R^0}_2 =  \frac{2r\dot{v} Qe^{01} -\left[4u\sqrt{1-kr^2} + r(4u^2-v^2)Q - 4r \ddot{Q}\right]e^{02} - 4ru\dot{Q}e^{12}}{4rQ} \, , \nonumber \\
   & &{R^1}_2 = \frac{2r u v Q^2 e^{01} - 4r\dot{u} Q^2 e^{02} + \left[ 4r\dot{Q}^2 + r v^2 Q^2+ 4kr - 4uQ\sqrt{1-kr^2}\right] e^{12}}{4rQ^2} \, .
  \nonumber
 \ea

\medskip

\noindent For the consistency of the equations, that is, for
keeping all $Q,u,v$ functions to be dependent only on the cosmic
time $t$ we have to assume $\Psi$ depending on both $t$ and $r$
such that $\Psi=\Psi(t,r) = \xi(t)/\sqrt{r}$. Thus the Dirac
equation (\ref{eq:DirEqn}) and its conjugate turn out to be
 \ba
     \dot{\xi} &=& \left[ -{\dot{Q}}/{Q} + ({3v}/{4} - m)\g_0 \right] \xi \, , \label{eq:DirEqn1}\\
     \dot{\overline{\xi}} &=&  \overline{\xi} \left[ -{\dot{Q}}/{Q} - ({3v}/{4} - m) \g_0 \right] \, .
 \ea
These results enable us to determine the Dirac energy-momentum 2-forms (\ref{eq:DirEnMom})
 \ba
   & & \tau_0 = 2\mathcal{S} [-u e^{01} + (2m-v)e^{12}] \, , \nonumber \\
   & & \tau_1 =- \frac{\mathcal{S}}{Q} [ 2\dot{Q}  e^{01} +  vQ e^{02}] \, , \nonumber \\
   & & \tau_2 =\frac{\mathcal{S}}{rQ} [rv Q e^{01} - 2r\dot{Q} e^{02} - 2 (ruQ + \sqrt{1-kr^2}) e^{12}] \, .
 \ea
Consequently we obtain the following class of solution to the
field equations (\ref{eq:FldEqCoframe}) and (\ref{eq:FldEqConnec})
 \ba
   u=0 \, , \quad v = \frac{ab-2c\l}{a^2 -2bc} \, , \quad
   \mathcal{S}=0 \, , \label{eq:solconstraint}\\
   Q(t) = \frac{1}{2h} \left[ e^{h (t + c_1)} + k e^{-h(t+c_1)} \right] \label{eq:Solution}
 \ea
where $c_1$ is a constant and $h^2 = [4(a^2 -2bc)(b^2 -a \l) - (ab
-2c\l)^2]/4(a^2-2bc)^2$. If $k<0$ and $c_1=(\ln{|k|})/(2h)$, the
space-time characterized by (\ref{eq:Solution}) which recalls the
inflationary solution of the General Relativity in four dimensions
has a singularity, i.e. $Q(0)=0$. Here we also notice that it must
be $a^2 - 2bc \neq 0$ for the solution, but even if $\l =0$ we
have a solution. The condition $a^2 -2bc \neq 0$  was shown to be
arising from the demands of a canonical constrained hamiltonian
analysis in \cite{banerjee2010}. For more readings one can consult
the equation (38) and the succeeding remarks of that work with the
replacement $a\rightarrow a$, $\alpha_3 \rightarrow 2c$ and
$\alpha_4 \rightarrow b$. However, a prescription to approach the
singular point $a^2 = 2bc $ and the canonical structure analysis
of the theory in the space of parameters  is discussed in
\cite{santamaria2011}.

\medskip

\noindent The next task is to solve the Dirac equation (\ref{eq:DirEqn1})
 \ba
   \left(\begin{array}{c}
     \dot{\xi_1}\\
     \dot{\xi_2}
   \end{array}\right)=
   \left(\begin{array}{cc}
   -\dot{Q}/Q &  3v/4-m \\
   -3v/4+m & -\dot{Q}/Q
   \end{array}\right)
   \left(\begin{array}{c}
   {\xi_1}\\ {\xi_2}
    \end{array}\right) \, .
 \ea
We can decouple these equations by virtue of a unitary transformation
 \ba
    \xi= U \rho \quad \mbox{with} \quad U=\frac{1}{\sqrt{2}} \left(\begin{array}{cc}
                                                               1 & 1 \\
                                                               i & -i
                                                               \end{array}\right) \, .
 \ea
Thus we arrive at the set of decoupled equations
 \ba
    \dot{\rho_1} = \left[-\dot{Q}/Q  + i(3v/4-m)\right] \rho_1 \, , \quad
    \dot{\rho_2} = \left[-\dot{Q}/Q  - i(3v/4-m)\right] \rho_2 \, .
 \ea
The solutions can be written as
 \ba
 \rho_1(t)=\frac{\rho_{+}}{Q(t)} \, e^{+i\phi(t)} \, , \quad  \rho_2(t)=\frac{\rho_{-}}{Q(t)} \, e^{-
 i\phi(t)} \label{eq:soluofdirac}
 \ea
where $\rho_{\pm}$ are the integration constants and $\phi =
(3v/4-m)t$ is the phase factor. Then we calculate the Dirac
condensate that is zero according to (\ref{eq:solconstraint}iii)
 \ba
    \mathcal{S} = \frac{i}{4} \overline{\Psi} \Psi = \frac{i}{4r} \overline{\xi} \xi
    = \frac{|\rho_{-}|^2 - |\rho_{+}|^2}{4rQ^2} =0 \, .
 \ea
Consequently it must be $ |\rho_{+}|^2 = |\rho_{-}|^2$.
Furthermore since $\mathcal{S} =0$, the Dirac spin 3-form is zero,
$\Sigma_{ab}=0$, via (\ref{eq:DirSpin}) as well.

\section{Concluding Remarks}\label{sec:conclusion}

In this work we considered the minimal coupling of a Dirac
particle to the MB-model of gravity in three dimensional
Riemann-Cartan spacetime. After computing the variational field
equations we considered the homogenous and isotropic metric, and
made an ansatz for torsion. Then we calculated the Dirac equation
which gave us the opportunity of obtaining the Dirac
energy-momentum 2-forms. Consequently we were at a position of
writing all field equations explicitly. Thus we obtained a family
of solution (\ref{eq:solconstraint}), (\ref{eq:Solution}) and
(\ref{eq:soluofdirac}) which is valid for $a^2 -2bc \neq 0$ and
even if $\l=0$.

\medskip

\noindent It is worthwhile to pay more attention to the result,
$\mathcal{S}=0$, given by the equation (\ref{eq:solconstraint}).
It may imply that even though the Dirac field has no effect on the
geometry, the later affects the time evolution of the spinor
field. Thus our solution may evoke the Einsteinian cosmological
solution satisfying perfect cosmological principle in which $\l
\neq 0$ is a crucial condition. However, our model and the
geometry are totally different from Einstein's theory and the
Riemann spacetime, respectively. In fact, since the presence of a
Dirac spinor modifies the geometry of space-time by affecting the
connection that determines the notion of parallel transport, it is
no longer possible to conclude that the geometry is totally
unaware of the presence of a Dirac spinor.

\medskip

\noindent Meanwhile, there are works in which the
Einstein-Cartan-Dirac (ECD) theory was investigated in four
dimensions, for example
\cite{seitz1985},\cite{dimakis1985},\cite{baekler1988} . Even they
were done in four dimensions, they are related to our work,
especially, with regards to $\mathcal{S}=0$. For clarifying the
relation we firstly notice that $\mathcal{S}=0$ means $
{}^*(\Sigma_{ab} \w {}^*\Sigma^{ab})=0$ equivalent to
$\Sigma_{abc}\Sigma^{abc}=0$ in the language of tensors with the
notation $\Sigma_{ab}:=\Sigma_{abc} {}^*e^c$. In four dimensions
because of the complexity of the ECD problem the authors of those
papers proceed with the following strategy. They firstly determine
orthonormal coframe $e^a$ by solving the Einstein equation (with
cosmological constant) in vacua. Then they give the constraint by
hand $\Sigma_{abc}\Sigma^{abc}=0$ for $\Psi$. Finally they compute
torsion and spinor through the field equations. For massless Dirac
spinors this constraint corresponds to the helicity state equation
$ \frac{1}{2}(1 \pm i\gamma_5)\Psi = \Psi$ \cite{seitz1985}. The
neutrinos are considered as massless fermions and correspondingly
have definite helicities according to the standard model.
Therefore, they may be treated in this context. Thus in the
Ref.\cite{dereli1981} even though the authors did not follow
directly the strategy outlined above, their results and
calculations are parallel to it. However, in three dimensions
because of relative simplicity one does not need to assume any
constraint like that from the beginning. We encounter with that as
a result of the equations. But this is not a generic result, that
is, for some other models and their solutions in (1+2)-dimensions
it may be $\Sigma_{abc}\Sigma^{abc} \approx \mathcal{S}^2 \neq 0$,
see e.g. \cite{dereli2010}.

\appendix

\section*{Appendix: Irreducible Decompositions}

In this section we give briefly the irreducible pieces of torsion,
contortion and curvature in three dimensions. Firstly torsion
which has nine components can be decomposed
 \ba
   \underbrace{T^a}_{\# 9} = \underbrace{ {}^{(1)}T^a}_{\# 5}
+ \underbrace{{}^{(2)}T^a}_{\# 3} + \underbrace{{}^{(3)}T^a}_{\#
1}
 \ea
where ${}^{(2)}T^a = - \frac{1}{2} (\iota_b T^b) \wedge e^a$, $
{}^{(3)}T^a = \frac{1}{3} \iota^a (e_b \w T^b)$ and ${}^{(1)}T^a =
T^a  - {}^{(2)}T^a - {}^{(3)}T^a$. In this section the notation
with the number under a brace is for the number of components of
that part. They have the properties, ${}^{(1)}T^a \w e_a =
{}^{(2)}T^a \w e_a =0$ and $\iota_a {}^{(1)}T^a = \iota_a
{}^{(3)}T^a = 0$. Our choice (\ref{eq:tors}) corresponds to
 \ba
    ^{(1)}T^{a}=0, \ \ ^{(2)}T^{a}= u\left(\begin{array}{c}
             e^{01}\\
                   0\\
             -e^{12} \\
           \end{array} \right),  \ \
 ^{(3)}T^{a}=v\left(\begin{array}{c}
             e^{12}\\
              e^{02}\\
             -e^{01} \\
           \end{array} \right)
 \ea
which means $4=3\oplus 1$. After the solution
(\ref{eq:solconstraint}) we are left only with ${}^{(3)}T^{a}$.

\medskip

\noindent Secondly one can decompose the contortion having nine
components
 \ba
   \underbrace{K_{ab}}_{\# 9}=\underbrace{{}^{(1)}K_{ab}}_{\# 5} +
 \underbrace{{}^{(2)}K_{ab}}_{\# 3} + \underbrace{{}^{(3)}K_{ab}}_{\# 1}
 \ea
where ${}^{(2)}K_{ab}=\frac{1}{2}[e_a \w (\iota^cK_{cb}) - e_b \w
(\iota^cK_{ca})]$, ${}^{(3)}K_{ab}=-\frac{1}{6}\iota_{ab}
(K_{cd}\w e^{cd})$ and ${}^{(1)}K_{ab} = K_{ab} - {}^{(2)}K_{ab} -
{}^{(3)}K_{ab}$. They have the properties
$\iota_a{}^{(1)}K^{ab}=\iota_a{}^{(3)}K^{ab}=0$ and $
{}^{(1)}K_{ab} \w e^{ab} = {}^{(2)}K_{ab} \w e^{ab}=0$. For our
case (\ref{eq:contor2}) we possess again ${}^{(1)}K_{ab}=0$, but
nonzero ${}^{(2)}K_{ab}$ and ${}^{(3)}K_{ab}$, i.e. $4=3\oplus1$.
Besides after the solution (\ref{eq:solconstraint}) only
${}^{(3)}K_{ab}$ survives.

\medskip

\noindent Finally one can split the curvature with nine components
  \ba
   \underbrace{R_{ab}}_{\# 9}=\underbrace{{}^{(1)}R_{ab}}_{\# 5} +
 \underbrace{{}^{(2)}R_{ab}}_{\# 3} + \underbrace{{}^{(3)}R_{ab}}_{\# 1}
 \ea
where ${}^{(2)}R_{ab}=\frac{1}{2}(e_a \w \iota_b - e_b \w
\iota_a)(e^c \w R_c)$, ${}^{(3)}R_{ab}=\frac{1}{6} R e_{ab}$ and
${}^{(1)}R_{ab} = R_{ab} - {}^{(2)}R_{ab} - {}^{(3)}R_{ab}$ with
$R_a = \iota^bR_{ba}$ and $R=\iota^aR_a$. They have the properties
$\iota_{ab}{}^{(1)}R^{ab}=\iota_{ab}{}^{(2)}R^{ab}=0$, $
{}^{(1)}R_{ab} \w e^b = {}^{(3)}R_{ab} \w e^b=0$ and $e_b \w
\iota_a {}^{(1)}R^{ab}=0$. For the solution
(\ref{eq:solconstraint}) although all three pieces of curvature
and of torsion are nonzero, since $DT^a =0$ the Ricci tensor is
symmetric.

\end{document}